\documentclass[12pt]{iopart}
\usepackage{amsfonts}
\usepackage{amssymb}
\usepackage{graphicx}
\usepackage{bm}
\usepackage{color}
\usepackage[colorlinks=true,citecolor=blue,linkcolor=blue,urlcolor=blue]{hyperref}
\usepackage{cite}
\usepackage{capt-of}
\usepackage{graphicx}
\usepackage{subcaption}
\usepackage{hyperref}

\begin{document}

\title{N-photon solutions to the two-qubit quantum Rabi model}

\author{ Qiang Lin$^{1}$, Junlong Tian$^{2}$, Pinghua Tang$^{1}$ and Jie Peng$^{1,*}$}

\address{$^1$ Hunan Key Laboratory for Micro-Nano Energy Materials and Devices and School of Physics and Optoelectronics, Xiangtan University, Hunan 411105, China}%
\address{$^2$ Department of electronic science, College of Big Data and Information Engineering, Guizhou University, Guiyang 550025, China}%
\eads{\mailto{jpeng@xtu.edu.cn}}\date{\today }
\begin{abstract}
  We studied the two-qubit quantum Rabi model and found its dark state solutions with at most N photons. One peculiar case presents when $N=3$, which has constant eigenenergy in the whole coupling regime and leads to level crossings within the  same parity subspace. We also discovered asymptotic solutions with at most $N=2i+3$ $(i=1,2,3,\dots)$ photons, and constant eigenenergy $N\hbar \omega$ when coupling $g$ becomes much larger than photon frequency $\omega$. Although generally all photon number states are involved in the two-qubit quantum Rabi model, such $N$-photon solutions exist and may have applications in quantum information processing with ultrastrong couplings. 
\end{abstract}

\vspace{2pc} \noindent Keywords: two-qubit quantum Rabi model,  N-photon solution, dark state, asymptotic solution,  level crossings.
\maketitle

\section{Introduction}
 Rabi introduced the semiclassical Rabi model in the 1930s to discuss the motion of an oriented atom with nuclear spin in a rapidly varying weak magnetic field \cite{I,II}.
  The quantum Rabi model (QRM) describing the interaction between a single-mode light field and a qubit has important applications in quantum optics \cite{2,3,4,5}, cavity quantum electrodynamics (QED) \cite{6}, circuit QED \cite{7}, quantum dots \cite{8}, among others.
  However, it is difficult to solve. Many approximations have been made, the most famous of which is the rotating wave approximation. Under this approximation, the Jaynes-Cummings (JC) model was proposed, which can be solved easily \cite{1}.
  However, this approximation does not hold in the ultrastrong coupling regime, which has already been realized experimentally. Evidence has been reported that the JC model is no longer applicable in these scenarios \cite{I3}. 
  Subsequent researches have extensively explored the full QRM Hamiltonian. Braak found an analytical solution for the QRM in Bargmann space in 2011 \cite{10,14} and Chen et al. \cite{15} retrived its solution using the Bogoliubov operator method.
  Subsequently, researches have been conducted on various generalized QRMs, including asymmetric QRMs \cite{14,15,16}, anisotropic QRMs \cite{17,18,19}, and two-photon QRMs \cite{20,21,22,23,24}.

  Meanwhile, the two-qubit QRM, which describes the coupling between two qubits and a single mode photon field, has been widely studied in \cite{26,27,28,29}. It has broad and important applications in cavity and circuit QED \cite{30,31,32,33}.
  Its solution generally consists of infinite photons \cite{27,35}. Interestingly, there exists a special dark state with at most one photon and constant eigenenergy \cite{36}.
  This enables the deterministic fast generation of W states \cite{37} and high-quality single photon sources \cite{38} via adiabatic evolution.
  Therefore, further exploration of dark state solutions with at most N photons may lead to other important applications in quantum information.
  
  In this paper, we studied the two-qubit QRM and found its exceptional dark state solutions with at most N photons for specific parameters. The necessary and sufficient condition for their existence is obtained by solving a secular equation. One such solution has constant eigenenergy in the whole coupling regime with $N=3$. It corresponds to a horizontal line in the spectrum with $E=3\hbar\omega$ in the spectrum, which will obviously cause level crossing within the parity subspace. We find an operator in the eigenenergy basis to label all the degeneracies. 
 Meanwhile, we found an asymptotic solution with at most $N=2i+3$ $(i=1,2,3,\dots)$ photons, and constant energy $N\hbar\omega$ when $g$ becomes much larger than $\omega$.
  Such solutions offer a method to obtain the exceptional solution of generalized QRMs with a photon number upper bound. Such solution may have applications in fast quantum information protocols with ultrastrong light-matter couplings.

\section{Two-qubit QRM and its N-photon solution}

The Hamiltonian of the two-qubit QRM reads ( $\hbar $ = 1)\cite{30,41}
\begin{equation}
        H_{tq}=\omega a^\dagger a+g_1\sigma_{1x}(a+a^\dagger)+g_2\sigma_{2x}(a+a^\dagger)+\Delta_1\sigma_{1z}+\Delta_2\sigma_{2z},
 \end{equation}
 where $a^\dagger$ and $a$ are the single mode photon creation and annihilation operators with frequency $\omega$, respectively. $\sigma_i\left(i=x,y,z\right)$ 
 are the Pauli matrices. $2\Delta_1,2\Delta_2$ are the energy level splittings. $g_1 $ and $g_2 $ are the qubit–photon coupling constants for the two qubits respectively. $H_{tq}$ commutes with a parity operator $\exp({i\pi a^\dag a})\sigma_{1z}\sigma_{2z}$.
 Supposing there is an eigenstate with at most N photons $|\Psi\rangle=c_{0,a}|0\uparrow\uparrow\rangle+c_{0,b}|0\downarrow\downarrow\rangle+c_{1,a}|1\uparrow\downarrow\rangle+c_{1,b}|1\downarrow\uparrow\rangle+\dots +c_{N,a}|N\uparrow\downarrow\rangle+c_{N,b}|N\downarrow\uparrow\rangle $ in th even parity subspace, where $N$ is odd, then the eigenenergy equation reads ($\omega$ is set to 1)
 \begin{equation}
    \fl
    \setlength{\arraycolsep}{3pt}
    \scriptsize 
    \left(\begin{array}{ccccccccc}0-E-\Delta_1-\Delta_2&0&g_1&g_2&0&\ldots
\\0&0-E+\Delta_2+\Delta_1&g_2&g_1&0&\ldots
\\\cdots&\cdots&\cdots&\cdots&\cdots&\cdots
\\0&\cdots& \sqrt{N}g_1 & \sqrt{N}g_2 & N-E+\Delta_1-\Delta_2 & 0
\\0&\cdots&  \sqrt{N}g_2 & \sqrt{N}g_1 & 0 & N-E+\Delta_2-\Delta_1 
\\0&\cdots&   0 & 0 & \sqrt{N+1}g_1& \sqrt{N+1}g_2
\\0&\cdots&   0 & 0 & \sqrt{N+1}g_2 & \sqrt{N+1}g_1\end{array}\right) \left(\begin{array}{ccccccccc}c_{0,a}\\c_{0,b}\\\cdots\\\cdots\\c_{N,a}\\c_{N,b}\end{array}\right)=0.
\label{3}
\end{equation}
There are more equations than variables, so generally it can not be solved. However, if the number of nonzero rows in the above matrix can be less than the number of columns after elementary row transformations, then there will be nontrivial solutions. 
The last two equations for $c_{N,a}$ and $c_{N,b}$ give $g_1=g_2=g$ and  $c_{N,a}=-c_{N,b}$. If $E=N$ simultaneously, then the coefficient matrix becomes a square one
\begin{equation}
    \setlength{\arraycolsep}{1pt}
    \small 
    \left(\begin{array}{ccccccccc}-N-\Delta_1-\Delta_2&0&g&g&0&\ldots\\0&-N+\Delta_2+\Delta_1&g&g&0&\ldots\\\ldots&\ldots&\ldots&\ldots&\ldots&\ldots\\0&\ldots&\sqrt Ng&\sqrt Ng&\Delta_1-\Delta_2&0\\0&\ldots&0&0&1&1\end{array}\right).
    \label{5}
\end{equation}
 If N=3, the matrix equation $(\ref{5})$ reduces to

 \begin{equation}
    \fl 
    \setlength{\arraycolsep}{3pt}
    \scriptsize 
\left(
\begin{array}{cccccccc}
-\Delta_1-\Delta_2-3 & 0 & g & g & 0 & 0 & 0 & 0 \\
0 & \Delta_2+\Delta_1-3 & g & g & 0 & 0 & 0 & 0 \\
g & g & \Delta_1-\Delta_2-2 & 0 & \sqrt{2}g & \sqrt{2}g & 0 & 0 \\
g & g & 0 & \Delta_2-\Delta_1-2 & \sqrt{2}g & \sqrt{2}g & 0 & 0 \\
0 & 0 & \sqrt{2}g & \sqrt{2}g & -\Delta_1-\Delta_2-1 & 0 & \sqrt{3}g & \sqrt{3}g \\
0 & 0 & \sqrt{2}g & \sqrt{2}g & 0 & \Delta_2+\Delta_1-1 & \sqrt{3}g & \sqrt{3}g \\
0 & 0 & 0 & 0 & \sqrt{3}g & \sqrt{3}g & \Delta_1-\Delta_2 & 0 \\
0 & 0 & 0 & 0 & 0 & 0 & 1 & 1
\end{array}
\right).
 \end{equation}
 \begin{table}
  \centering
  \caption{The conditions for $N$-photon solutions when N equal to  2, 3, and 4, where ${{{a}=\Delta_{1}-\Delta_{2}}},{{{b}=\Delta_{1}+\Delta_{2}}}$.}
  \label{tab:my_table}
  
  \begin{tabular}{lll} 
      \hline
      N &     $\rm{Condition}$ \\
      \hline
      2 & $g=\frac{\sqrt{-4+b^2}\sqrt{-1+a^2}}{2\sqrt2}$ \\

     3 & $ g\sqrt{168-40b^2}=\sqrt{-9+10b^2-b^4}\sqrt{-4+a^2}$  \\
  
      4 & $g=\frac{\sqrt{184-56a^2+b^2\left(-13+5a^2\right)+\sqrt{b^4\left(97-50a^2+17a^4\right)-16\mathrm{~b}^2\left(209-106a^2+25a^4\right)+64\left(457-242a^2+41a^4\right)}}}{4\sqrt{6}}$ \\

  \end{tabular}

\end{table}

 \begin{figure}
  \centering
  \includegraphics[width=0.7\textwidth]{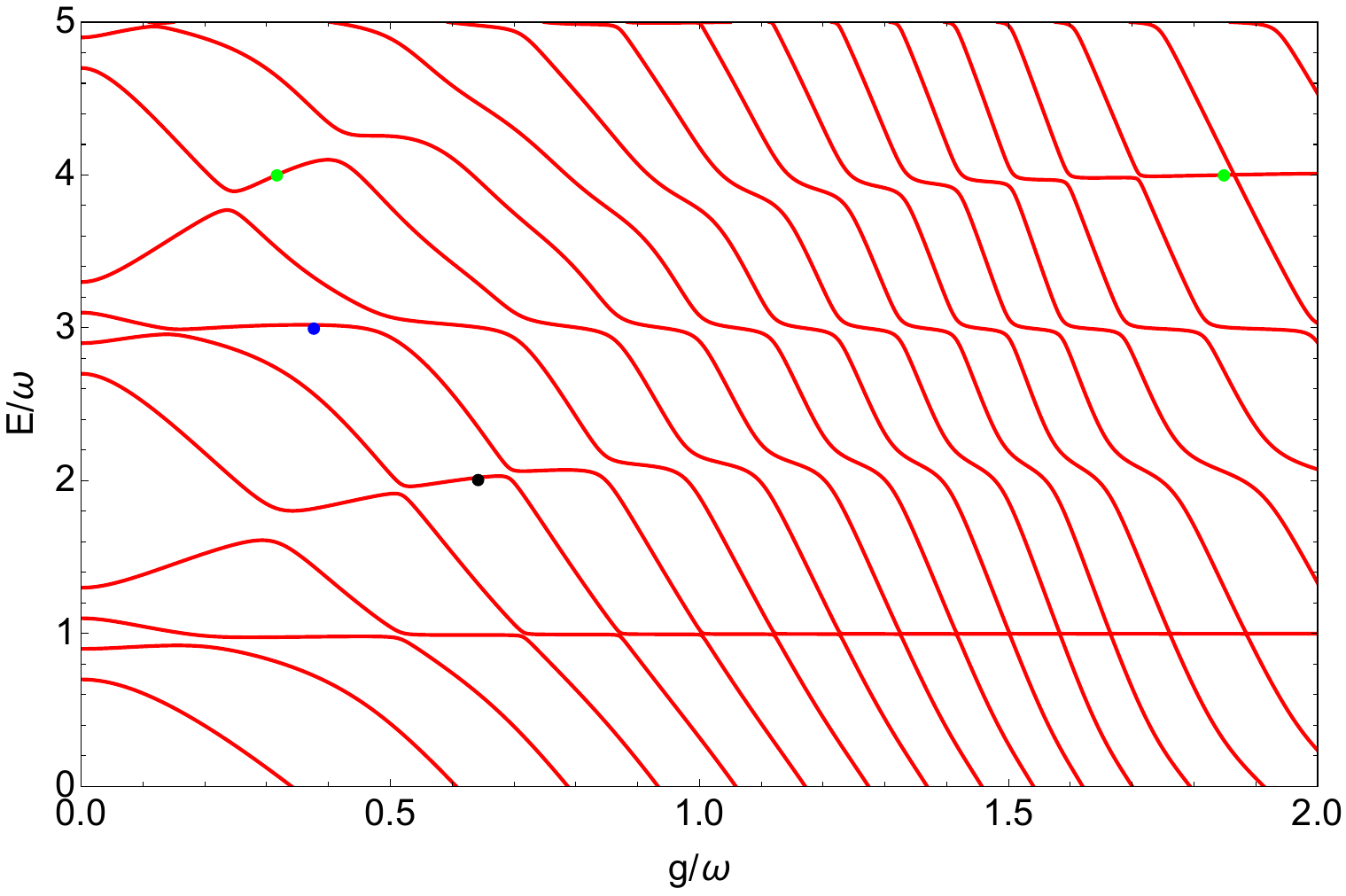}
  \caption{The numerical spectrum of the two-qubit QRM with even parity. $\Delta_1=0.6 ,\Delta_2=0.3,g_1=g_2, \omega=1, 0 \leqslant g = g_1=g_2 \leqslant 2.$ The black, blue, and green dots correspond to dark states with at most 2, 3, 4 photons, respectively.}
  \label{1}
\end{figure} 
The secular equation reads
$$\fl\left(\Delta_1-\Delta_2\right)^2\left[-36+168g^2+\Delta_1^6+2\Delta_1^5\Delta_2+\left(49-40g^2\right)\Delta_2^2-14\Delta_2^4~~~~~~~~~\right.$$
$$\fl+\Delta_2^6-4\Delta_1^3\Delta_2\left(4+\Delta_2^2\right)-\Delta_1^4\left(14+\Delta_2^2\right)+2\Delta_1\Delta_2\left(31-40g^2-8\Delta_2^2+\Delta_2^4\right)$$
\begin{equation}
    \fl\left.-\Delta_1^2\left(-49+40g^2+4\Delta_2^2+\Delta_2^4\right)\right]=0,
\end{equation}
which reduces to 
\begin{equation}
    g\sqrt{168-40b^2}=\sqrt{-9+10b^2-b^4}\sqrt{-4+a^2},
    \label{6}
\end{equation}
where ${{{a}=\Delta_{1}-\Delta_{2}}}, {{{b}=\Delta_{1}+\Delta_{2}}}$. Substituting equation (\ref{6}) into equation (\ref{3}), we can obtain the corresponding eigenstate. Using the same approach, we can obtain the N-photon solutions. We display the existence conditions for $N=2,3,4$ in table \ref{tab:my_table}. We choose $\Delta_1=0.6, \Delta_2=0.3$, and depict the numerical spectrum of the two-qubit QRM in figure \ref{1}. The black, blue, and green dots correspond to dark states solutions with at most 2, 3, and 4 photons. Their energies are $2\hbar\omega, 3\hbar\omega$ and $4\hbar\omega$, respectively. It is evident that these solutions are discrete points, similar to the Juddian solution of the QRM \cite{42}.

\section{The special dark state for N=3}
\begin{figure}
    \centering
    \includegraphics[width=0.6\textwidth]{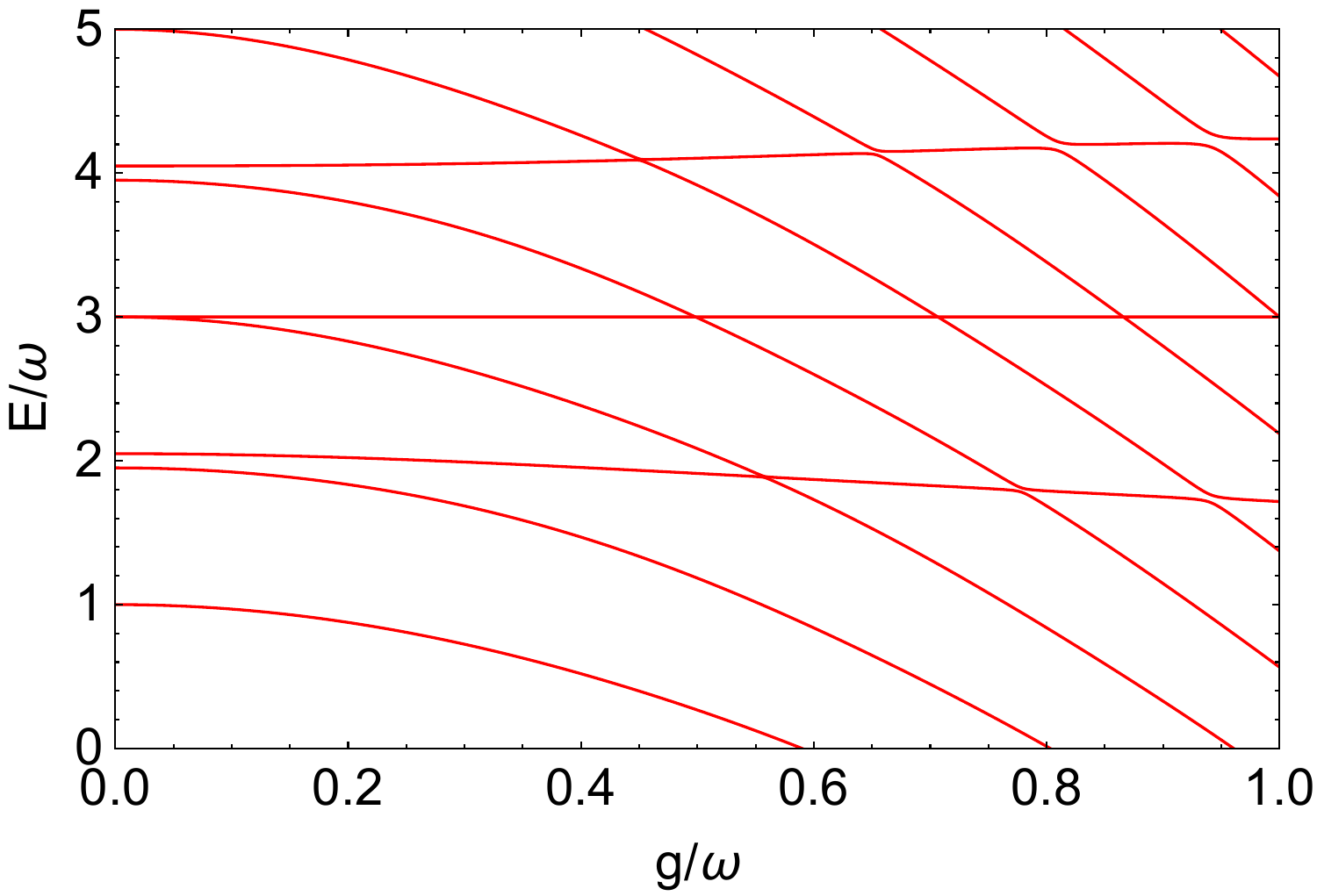}
    \caption{The numerical spectrum of two-qubit QRM in even parity with $\Delta_1=\frac{1}{2}\left(-2+\sqrt{\frac{21}{5}}\right) ,\Delta_2=\frac{1}{2}\left(2+\sqrt{\frac{21}{5}}\right),g_1=g_2, \omega=1, 0 \leqslant g = g_1=g_2 \leqslant 1.$}
    \label{fig:your-figure}
  \end{figure}

N-photon dark states generally corresponds to isolated points in the spectrum, so their existence is depdent on fine tuning with respect to couplings and qubit energies. 
It is interesting to explore solutions existing for arbitrary couplings at certain $\Delta_{1,2}$, just like the special dark state with $N=1$ \cite{27}.
This can be done when $N=3$, as can be seen in table \ref{tab:my_table}. If both sides of the equation (\ref{6}) are equal to 0, then the existence condition will be independent of g, which reads
\begin{equation}
    b=\sqrt{\frac{21}{5}}  ,  a=\pm2.
\end{equation}
So \begin{equation}\Delta_1=\frac{1}{2}\left(-2+\sqrt{\frac{21}{5}}\right),~ \Delta_2=\frac{1}{2}\left(2+\sqrt{\frac{21}{5}}\right),\label{8}\end{equation}
or 
\begin{equation}\Delta_1=\frac{1}{2}\left(2+\sqrt{\frac{21}{5}}\right), ~  \Delta_2=\frac{1}{2}\left(-2+\sqrt{\frac{21}{5}}\right).\end{equation}
Under the condition equation (\ref{8}), we can obtain one special dark state
$$
|{\psi_{e}}\rangle =\frac{270\sqrt{2}-26\sqrt{210}}{165\sqrt{3}g-45\sqrt{35}g}|0,g,g\rangle+\frac{8\sqrt{2}}{15\sqrt{3}
\mathrm{~g-3~\sqrt{35}~g}}|0,e,e\rangle$$ $$+\frac{\sqrt{2}\left(-5+\sqrt
{105}\right)\left(-3+\sqrt{105}\right)}{15\left(5\sqrt{3}-\sqrt{35}\right)\mathrm{~g}^2}|1,e,g\rangle$$ $$+
\frac{4\left(-9+\sqrt{105}\right)}{3\left(5\sqrt{3}-\sqrt{35}\right)g}|2,g,g\rangle $$\begin{equation}-\frac
{2\left(-3+\sqrt{105}\right)}{3\left(5\sqrt{3}-\sqrt{35}\right)g}|2,e,e\rangle-|3,e,g\rangle+|3,g,e\rangle\bigg.
\label{bo1}.
\end{equation}
The other solution can be obtained by swapping the qubits in the above solution. 
Similarly, for odd parity, the existence condition reads  
\begin{equation}
    {\fl -36b^2+40a^2b^2-4a^4b^2+9b^4-10a^2b^4+a^4b^4+168b^2g^2-40a^2b^2g^2=0},
\end{equation}
which gives
\begin{equation}
    b=2 ,  a=\pm\sqrt{\frac{21}{5}}.
\end{equation}
Because 
$ |\Delta_1 + \Delta_2| \geq |\Delta_1 - \Delta_2| $, this solution is impossible and such special dark state only exists in even parity. We will demonstate it only exists for N=3 in the following sections. Such solution exists in the whole coupling regime with constant energy $E=3\hbar\omega$, which apparently cause level crossing within the same parity subspace, as shown in the numerical spectrum figure $\ref{fig:your-figure}$.
We can use an operator which commutes with $H$ in the eigenergy basis to label the degeneracies, where  $ \hat{S} = \sum_{i,j} s_i | \psi_{i,j} \rangle \langle \psi_{i,j} | $ and $ |\psi_{i,j} \rangle $ is the j-th eigenstate of $\hat{S} $ with eigenvalue $ s_i $ \cite{39}.
  Here level crossings occur only between $|\psi_{e}\rangle$ and other energy levels. Therefore, we can express $\hat{S}=|\psi_{e}\rangle\langle\psi_{e}|+f(\Delta_{1,2},g)\sum_{\psi\neq\psi_{e}}|\psi\rangle\langle\psi|.$ Choosing $f(\Delta_{1,2},g)=0$ , we can obtain an analytical form of  $\hat{S} $ \cite{40}.   

  \section{Searching for special dark state with constant energy $E=N>3$}
  
  As can be seen in the last section, if the coefficients of $g^n$ in the determinant vanish, then the existence condition will be independent of couplings. Now we explore whether such solutions exist when $N>3$ by studying the square matrix equation (\ref{5}) for even $N$
 \begin{equation}
  \fl 
  \setlength{\arraycolsep}{3pt}
  \scriptsize 
\left(
\begin{array}{cccccccccccc}
-b-N & 0 & g & g & 0 & 0 & 0 & 0 & 0 & 0 & \dots \\
0 & b-N & g & g & 0 & 0 & 0 & 0 & 0 & 0 & \dots  \\
g & g & a-N+1 & 0 & \sqrt{2}g & \sqrt{2}g & 0 & 0 & 0 & 0 & \dots \\
g & g & 0 &-a-N+1 & \sqrt{2}g & \sqrt{2}g & 0 & 0 & 0 & 0 & \dots \\
0 & 0 & \sqrt{2}g & \sqrt{2}g & -b-N+2 & 0 & \sqrt{3}g & \sqrt{3}g & 0 & 0 & \dots\\
0 & 0 & \sqrt{2}g & \sqrt{2}g & 0 & b-N+2 & \sqrt{3}g & \sqrt{3}g & 0 & 0 & \dots\\
\dots & \dots & \dots & \dots & \dots & \dots & \dots & \dots & \dots & \dots \\
\dots & 0 & 0 & \sqrt{N-2}g & \sqrt{N-2}g & -b+2 & 0 & \sqrt{N-1}g & \sqrt{N-1}g & 0 & 0\\
\dots & 0 & 0 & \sqrt{N-2}g & \sqrt{N-2}g & 0 & b+2 & \sqrt{N-1}g & \sqrt{N-1}g & 0 & 0\\
\dots & 0 & 0 & 0 & 0 & \sqrt{N-1}g & \sqrt{N-1}g & a+1 & 0 & \sqrt{N}g & \sqrt{N}g\\
\dots & 0 & 0 & 0 & 0 & \sqrt{N-1}g & \sqrt{N-1}g & 0 & -a+1 & \sqrt{N}g & \sqrt{N}g\\
\dots & 0 & 0 & 0 & 0 & 0 & 0 & \sqrt{N}g & \sqrt{N}g & -b & 0 \\
\dots & 0 & 0 & 0 & 0 & 0 & 0 & 0 & 0 & b & b
\end{array}
\right).
\end{equation}
After elementary transformation, it reduces to
\begin{equation}
  \fl 
  \setlength{\arraycolsep}{3pt}
  \scriptsize 
\left(
\begin{array}{cccccccccccc}
-2N & N-b & 0 & 0 & 0 & 0 & 0 & 0 & 0 & 0 & \dots \\
N-b& b-N & 0 & g & 0 & 0 & 0 & 0 & 0 & 0 & \dots  \\
0 & 0 & -2N+2 & a+N-1 & 0 & 0 & 0 & 0 & 0 & 0 & \dots \\
0 & g & a+N-1 &-a-N+1 & 0 & \sqrt{2}g & 0 & 0 & 0 & 0 & \dots \\
0 & 0 & 0 & 0 & -2N+4 & -b+N-2 & 0 & 0 & 0 & 0 & \dots\\
0 & 0 & 0 & \sqrt{2}g & -b+N-2& b-N+2 & 0 & \sqrt{3}g & 0 & 0 & \dots\\
\dots & \dots & \dots & \dots & \dots & \dots & \dots & \dots & \dots & \dots \\
\dots & 0 & 0 & 0 & 0 & -4 & -b-2 & 0 & 0 & 0 & 0\\
\dots & 0 & 0 & 0 & \sqrt{N-2}g & -b-2 & b+2 & 0 & \sqrt{N-1}g & 0 & 0\\
\dots & 0 & 0 & 0 & 0 & 0 & 0 & -2 & a-1 & 0 & 0\\
\dots & 0 & 0 & 0 & 0 & 0 & \sqrt{N-1}g & a-1 & -a+1 & 0 & \sqrt{N}g\\
\dots & 0 & 0 & 0 & 0 & 0 & 0 & 0 & \sqrt{N}g & -b & 0 \\
\dots & 0 & 0 & 0 & 0 & 0 & 0 & 0 & 0 & 0 & b
\end{array}
\right).
\end{equation}
By analyzing the determinant of the above matrix, we find the term with highest power in $g$ reads
$$
 (-1)^{\frac{N}{2}} ({\prod_{i=1}^{\frac{N}{2}}\sqrt{N-2i+1}})g^{N}*
 \setlength{\arraycolsep}{3pt}
 \small 
 \rm{Det}
\left(
\begin{array}{cccccccccccc}
-2N & 0 & 0 & 0 & 0 & 0 & 0 & 0 & \dots \\
0 & -2N+2 & 0 & 0 & 0 & 0 & 0 &0& \dots \\
0 & 0 & -2N+4 & 0 & 0 & 0 & 0 & 0 & \dots \\
\dots & \dots & \dots & \dots & \dots & \dots & \dots & \dots & \dots\\
\dots & 0 & 0 & 0 & 0 & -4 & 0 & 0 & 0\\
\dots & 0 & 0 & 0 & 0 & 0 & -2 & 0 & 0\\
\dots & 0 & 0 & 0 & 0 & 0 & 0 & -b & 0\\
\dots & 0 & 0 & 0 & 0 & 0 & 0 & 0 & b
\end{array}
\right)
$$
\begin{equation}
  =
 (-1)^{\frac{N}{2}+1} b^{2}({\prod_{i=1}^{\frac{N}{2}}\sqrt{N-2i+1}})(\prod_{i=0}^{N-1} (-2N+2i))g^{N}.
\end{equation}
We clearly see the solution can not be indepdent of $g$ since $b\neq0$. When $N$ is odd, we find that the coefficient of highest power in $g$ is a function of $a$ and $b$ by using the same method. Therefore, it is still possible to find a solution independent of $g$ when $N$ is odd.
\section{The asymptotic solution for $N>3$}
 For N=5, the secular equation reads
      $$\fl { -14400a^2+4500a^4-225a^6+16576a^2b^2-5180a^4b^2+259a^6b^2-2240a^2b^4}$$
      $$ {\fl+700a^4b^4-35a^6b^4+64a^2b^6-20a^4b^6+a^6b^6+
      156480a^2g^2-12120a^4g^2~~~}$$ $$\fl{-60544a^2b^2g^2+6256a^4b^2g^2+2368a^2b^4g^2-280a^4b^4g^2-105600a^2g^4~~~~~~}$$\begin{equation}\fl{-105600a^2g^4+11392a^2b^2g^4=0}.\end{equation}
      $g$-independent solutions exist when the coefficient of $g^n$ are $0$, which leads to
        $$\fl{ -14400a^2+4500a^4-225a^6+16576a^2b^2-5180a^4b^2+259a^6b^2-2240a^2b^4}$$\begin{equation}\fl{+700a^4b^4-35a^6b^4+64a^2b^6-20a^4b^6+a^6b^6=0},\end{equation}
       \begin{equation}{  \fl -280a^4b^4+156480a^2-12120a^4-60544a^2b^2+6256a^4b^2+ 2368 a^2b^4=0}\label{2},\end{equation} 
        \begin{equation}\fl{-105600a^2+11392a^2b^2=0}\label{4}.\end{equation}
        This equation set has two variables and three equations, so that it has no solution normally. There are more equations when $N>5$ but the variables are still just $a$ and $b$, so generally it is impossible to obtain a $g$-independent 
        solution with $N>3$. We only need to change $a$ into $-b$ and $b$ into $-a$ for odd parity, so the same conclusion can be drawn. To conclude, g-independent solutions only exist when $N=1$ \cite{27} or $N=3$.

        However, asymptotic $g$-independent solution for $N>3$ can be obtained when  $g\gg \omega$. By setting the coefficients of two highest powers of $g$ to $0$, we can obtain solutions $\Delta_1=\Delta_{10}$, $\Delta_2=\Delta_{20}$.
        In this case, the determinant of the square matrix is not equal to 0, but we can prove that it tends to 0 when $g$ is relatively large.
        We denote the determinant as $\mathsf{F}(\Delta_1,\Delta_2,g)=\sum_{k=0}^{N}A_{k}\left(\Delta_1,\Delta_2\right)g^{2\left\lfloor k/2\right\rfloor}$ where $A_{k}\left(\Delta_1,\Delta_2\right)$ is the coefficient of $ g^{2\left\lfloor k/2\right\rfloor}$ and $\lfloor \rfloor$ means round down, so that 
        $\mathsf{F}(\Delta_{10},\Delta_{20},g)=\sum_{k=0}^{N-2}A_{k}\left(\Delta_{10},\Delta_{20}\right)g^{2\lfloor k/2\rfloor}$.
        To prove that as g tends to infinity, the solution to the secular equation will tend to $\Delta_{10},\Delta_{20} $, we perform the first-order Taylor expansion for $\Delta_{1}$ around $\Delta_{10}$, which gives
        
        \begin{figure}
          \centering
          \begin{subfigure}[c]{0.7\textwidth}
            \centering
            \includegraphics[width=\textwidth]{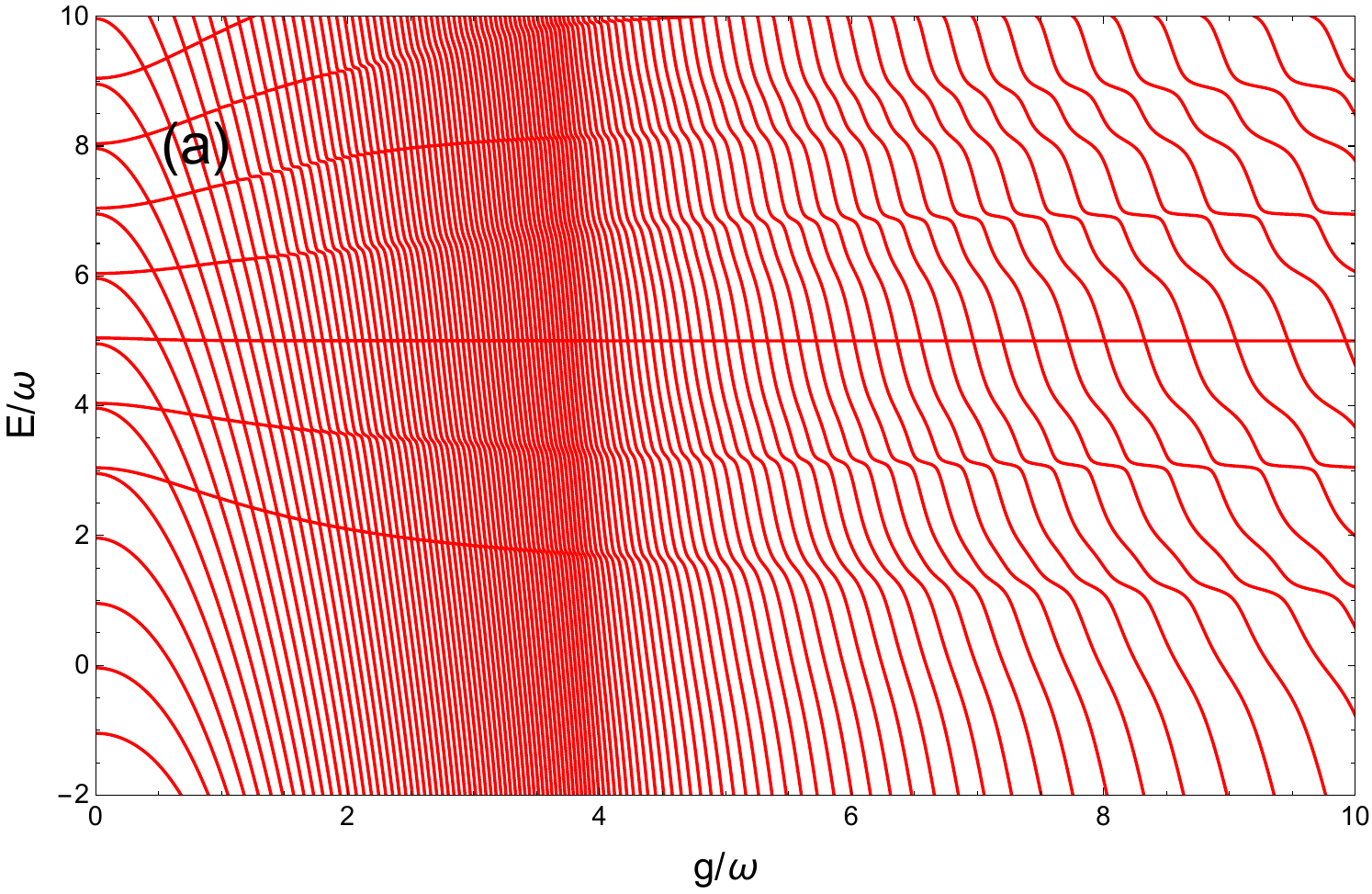}
            
            \label{fig:a}
          \end{subfigure}
          
          \begin{subfigure}[c]{0.7\textwidth}
            \centering
            \includegraphics[width=\textwidth]{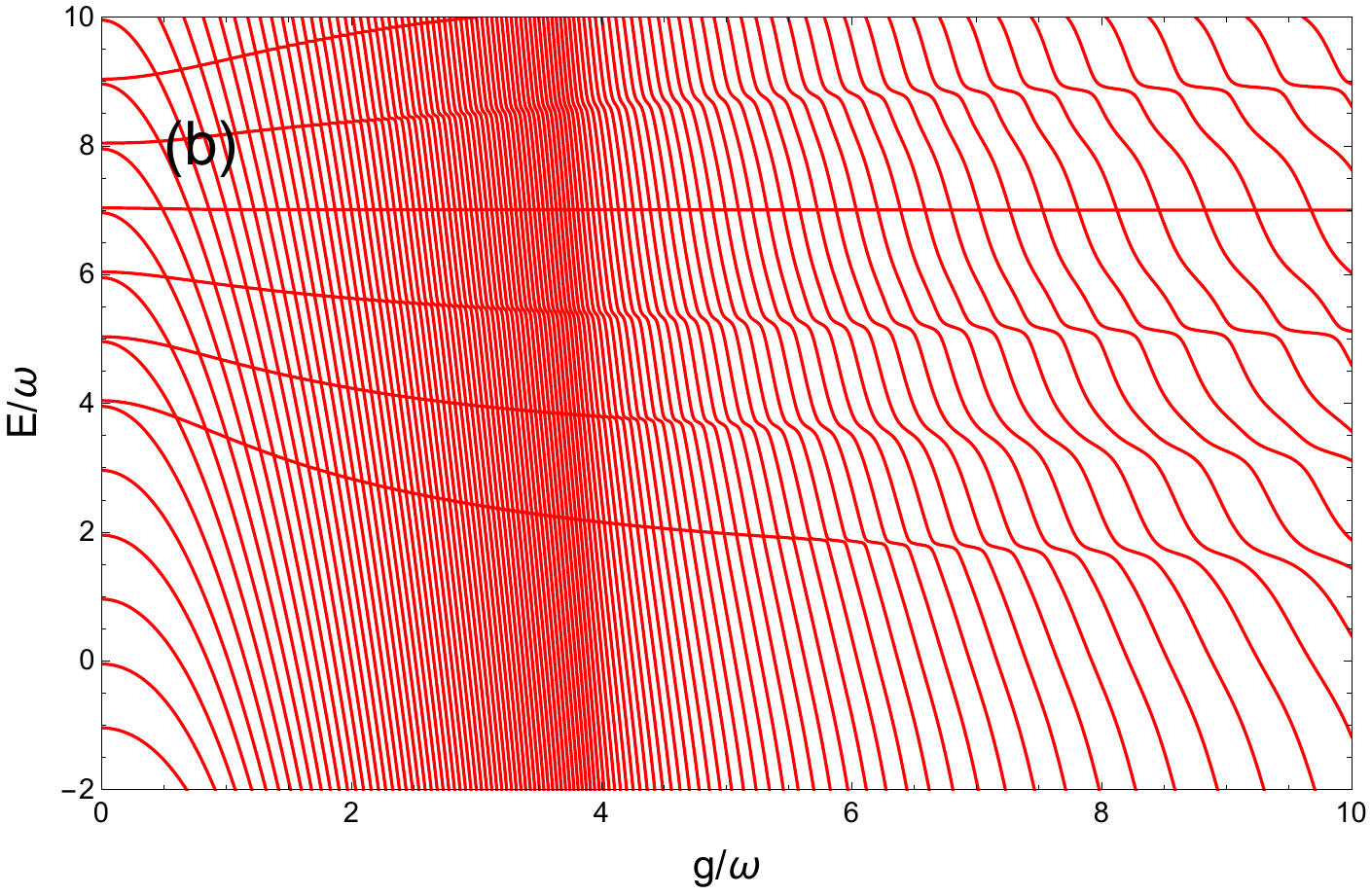}
            
            \label{fig:b}
          \end{subfigure}
          
          \caption{ (a) The numerical spectrum of the two-qubit QRM. An asymptotic solutions for N=5 in even parity with $\Delta_{1}=\frac{\sqrt{\frac{6487}{703}}+5\sqrt{\frac{33}{89}}}{2},\Delta_{2}=\frac{-\sqrt{\frac{6487}{703}}+5\sqrt{\frac{33}{89}}}{2}, g_1=g_2, \omega=1, 0 \leqslant g = g_1=g_2 \leqslant 10. $ (b) The numerical spectrum of asymptotic solutions for N=7 in even parity with $\Delta_1=4.041154, \Delta_2=0.003458, g_1=g_2, \omega=1, 0 \leqslant g = g_1=g_2 \leqslant 10.$  }
          \label{fig:ab}
        \end{figure}
        \begin{figure}
          \centering
          \includegraphics[width=0.79\textwidth]{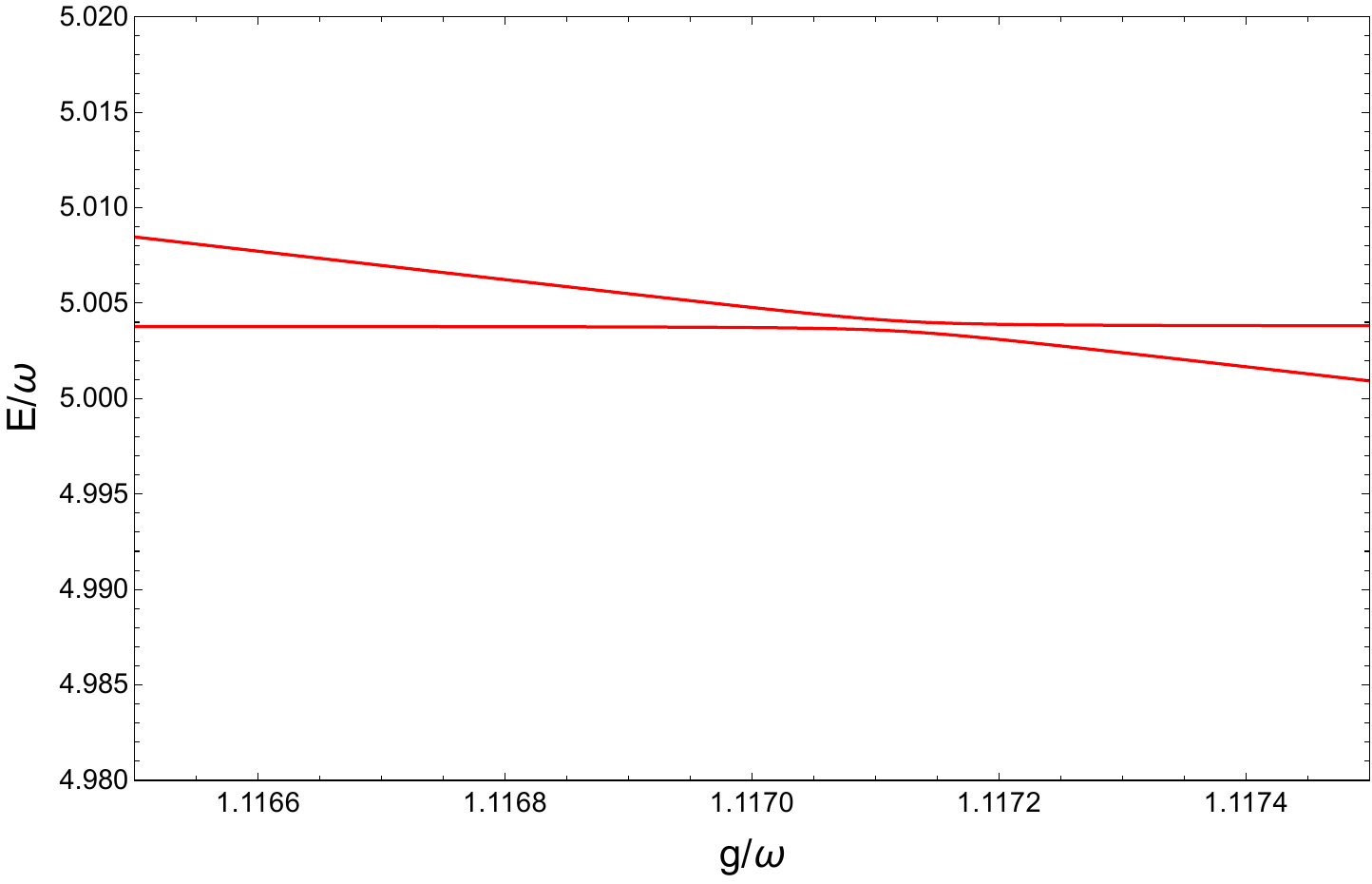}
          \caption{Zoom-in of figure \ref{fig:ab} around $g/\omega=1.1170$ and $E=5\hbar\omega$.  }
          \label{fig:avoidance}
        \end{figure}
        \begin{figure}
          \centering
          \includegraphics[width=0.79\textwidth]{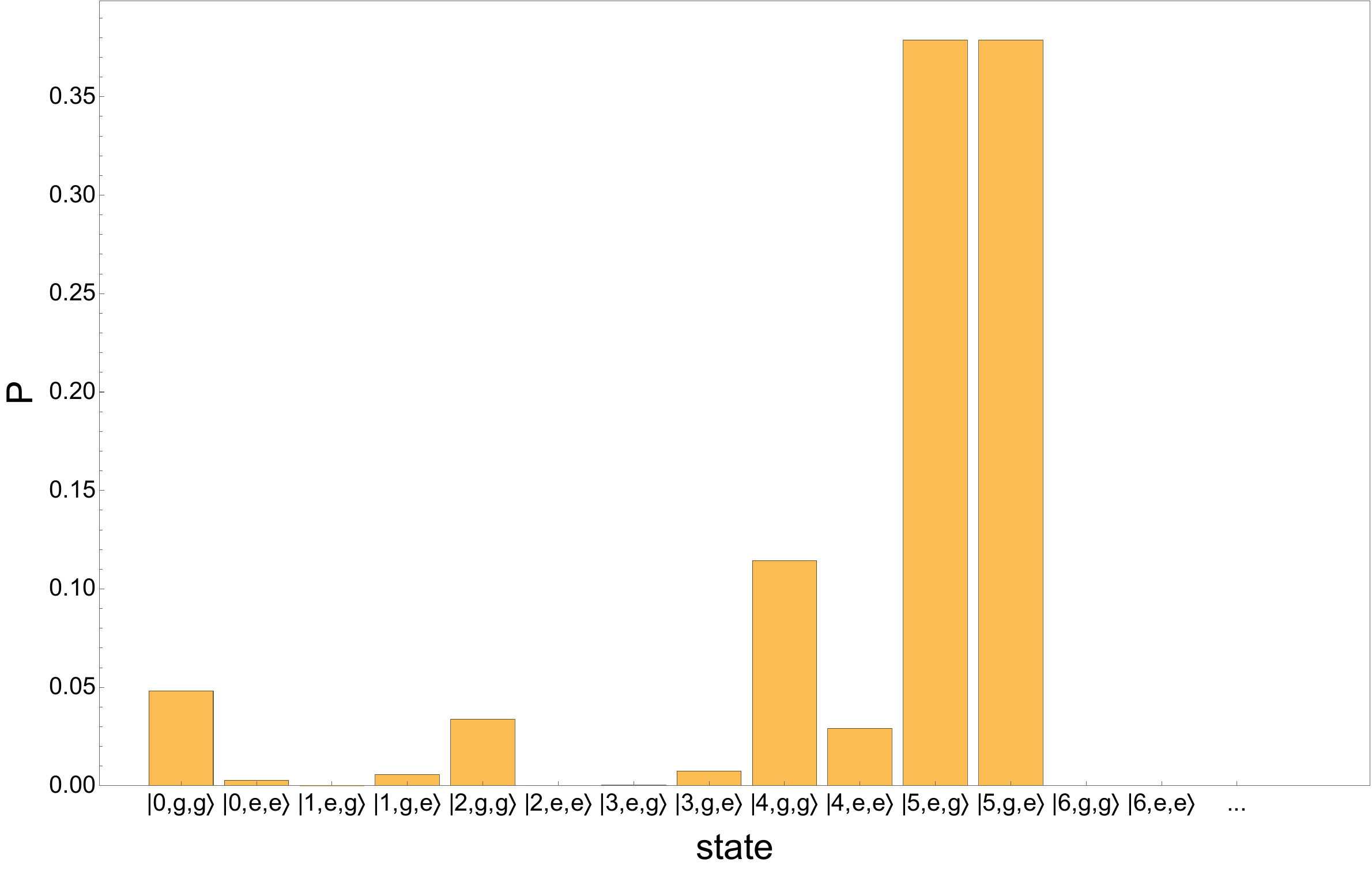}
          \caption{Population of states for the asymptotic solution with $E=N=5\hbar\omega$ when $\Delta_{1}=\frac{\sqrt{\frac{6487}{703}}+5\sqrt{\frac{33}{89}}}{2},\Delta_{2}=\frac{-\sqrt{\frac{6487}{703}}+5\sqrt{\frac{33}{89}}}{2}$, and $g=5\omega$.}
          \label{fig:5state}
        \end{figure}
        \begin{equation}
            \mathsf{F}(\Delta_1,\Delta_2,g) = \mathsf{F}(\Delta_{10},\Delta_{20},g) + \mathsf{F}^{\prime}(\Delta_{10},\Delta_{20},g)\delta \Delta_1 = 0,
        \end{equation}
        \begin{equation}
            \delta \Delta_1 = \frac{-\mathsf{F}(\Delta_{10},\Delta_{20},g)}{\mathsf{F}^{\prime}(\Delta_{10},\Delta_{20},g)} = -\frac{\sum_{k=0}^{N-2}A_{k}\left(\Delta_{10},\Delta_{20}\right)g^{2\lfloor k/2\rfloor}}{\sum_{k=0}^{N}B_{k}\left(\Delta_{10},\Delta_{20}\right)g^{2\lfloor k/2\rfloor}},
        \end{equation}
        where $B_{k}\left(\Delta_{10}, \Delta_{20}\right)$ is the coefficient of $g^{2\lfloor k/2\rfloor}$ in $\mathsf{F}^{\prime}(\Delta_{10},\Delta_{20},g)$. 
        Obviously, as $g$ goes to infinity, $\delta \Delta_1$ tends to 0,  indicating that $\Delta_{10}$ and $\Delta_{20}$ are the solution of the secular equation. So asymptotic solution with constant energy $E=N$ exists when $g$ goes much larger than $\omega$.
       It corresponds to a horizontal line in the spectrum with $E=N$ when $g \gg \omega$, as shown in figure \ref{fig:ab}. It seems they cause level crossings as the special dark state with $N=3$, however, they are actually narrow avoided crossings, as shown in figure \ref{fig:avoidance}. The photon number of the asymptotic solution is almost limited to $5$ when $g=5\omega$, according to the numerical results shown in figure \ref{fig:5state}.

\section{ Conclusion}
The solutions to the two-qubit QRM with photon number bounded from above at arbitrary $N$ are found, including special dark states with constant eigenenergy in the whole coupling regime when $N=3$.
 We also discover asymptotic solutions with constant energies $E=N=5, 7, 9,\ldots$ as $g$ becomes much larger than $\omega$. Such solutions deepen the understanding of generalized QRMs and may have 
 applications in fast quantum information protocols with ultrastrong light-matter couplings. 

\ack{}
This work was supported by the Scientific Research Fund of Hunan Provincial Education
 Department (Grants No. 23A0135, 21B0136),  Natural
 Science Foundation of Hunan Province, China (Grants Nos.
 2022JJ30556, 2023JJ30596, and 2023JJ30588), National Natural Science Foundation of China (Grants No. 11704320).

\end{document}